\documentclass[final]{aipproc}
\layoutstyle{8x11double}
\usepackage{amssymb}

\newcommand{ \be}{\begin{equation}}
\newcommand{ \ee}{\end{equation}}
\newcommand{ \bea}{\begin{eqnarray}}
\newcommand{ \eea}{\end{eqnarray}}
\newcommand{ \mysmall}[1]{\scriptscriptstyle #1} 
\newcommand{ \amu}{a_{\mu}}
\newcommand{ \mw}{M_{\mysmall{W}}}
\newcommand{ \mz}{M_{\mysmall{Z}}}
\newcommand{ \mh}{M_{\mysmall{H}}}
\newcommand{ \mhUB}{M_{\mysmall{H}}^{\mysmall \rm UB}}
\newcommand{ \mhLB}{M_{\mysmall{H}}^{\mysmall \rm LB}}
\newcommand{ \mt}{M_{t}}
\newcommand{ \mpi}{m_{\pi}}
\newcommand{ \seff}{\sin^2 \!\theta_{\rm eff}^{\rm lept}}
\newcommand{ \eq}[1]{Eq.~(\ref{eq:#1})}

\newcommand{ \gmt}  {$g$$-$$2$~}
\newcommand{ \dafive}{\Delta \alpha_{\rm had}^{\mysmall{(5)}}(\mz)}

\begin{document}

\title{The muon g-2 discrepancy: errors or new physics?}


\author{M.~Passera}{address={Istituto Nazionale Fisica Nucleare,
Sezione di Padova, I-35131, Padova, Italy}}

\author{W.~J.~Marciano}{address={Brookhaven National Laboratory, 
Upton, New York 11973, USA}}

\author{A.~Sirlin}{address={Department of Physics, New York University, 
10003 New York NY, USA}}

\begin{abstract}
  After a brief review of the muon \gmt status, we discuss hypothetical errors   in the Standard Model prediction that could explain the present discrepancy   with the experimental value. None of them looks likely. In particular, an   hypothetical increase of the hadroproduction cross section in low-energy   $e^+e^-$ collisions could bridge the muon \gmt discrepancy, but is shown to   be unlikely in view of current experimental error estimates. If,   nonetheless, this turns out to be the explanation of the discrepancy, then   the 95\% CL upper bound on the Higgs boson mass is reduced to about 130~GeV   which, in conjunction with the experimental 114.4~GeV 95\% CL lower bound,   leaves a narrow window for the mass of this fundamental particle.
\end{abstract}
\maketitle

\section{Introduction}

The anomalous magnetic moment of the muon, $a_{\mu}$, is one of the most interesting observables in particle physics.  Indeed, as each sector of the Standard Model ({\small SM}) contributes in a significant way to its theoretical prediction, the precise $a_{\mu}$ measurement by the E821 experiment at Brookhaven~\cite{bnl} allows us to test the entire {\small SM} and scrutinize viable ``new physics'' appendages to this theory~\cite{NP}.

The {\small SM} prediction of the muon \gmt is conveniently split into {\small   QED}, electroweak ({\small EW}) and hadronic (leading- and higher-order) contributions:
$
    a_{\mu}^{\mysmall \rm SM} = 
         a_{\mu}^{\mysmall \rm QED} \!+\!
         a_{\mu}^{\mysmall \rm EW}  \!+\!
         a_{\mu}^{\mbox{$\scriptscriptstyle{\rm HLO}$}} \!+\!
         a_{\mu}^{\mbox{$\scriptscriptstyle{\rm HHO}$}}.
$  
The {\small QED} prediction, computed up to four (and estimated at
five) loops, currently stands at
$a_{\mu}^{\mysmall \rm QED} = 116584718.10(16)
\!\times\! 10^{-11}$\cite{QED}, 
while the {\small EW} effects provide
$a_{\mu}^{\mysmall \rm EW} = 154(2) \!\times\! 10^{-11}$\cite{EW}.
The latest calculations of the hadronic leading-order contribution, via the hadronic $e^+ e^-$ annihilation data, are in good agreement:
$ a_{\mu}^{\mbox{$\scriptscriptstyle{\rm HLO}$}} = 
 6909(44) \!\times\! 10^{-11}$\cite{DE07},
$6894(46) \!\times\! 10^{-11}$\cite{HMNT06,TeubnerSUSY08},
$6921(56) \!\times\! 10^{-11}$\cite{Jeger06}, and
$6944(49) \!\times\! 10^{-11}$\cite{TY05}. 
The higher-order hadronic term is further divided into two parts:
$
     \amu^{\mbox{$\scriptscriptstyle{\rm HHO}$}}=
     \amu^{\mbox{$\scriptscriptstyle{\rm HHO}$}}(\mbox{vp})+
     \amu^{\mbox{$\scriptscriptstyle{\rm HHO}$}}(\mbox{lbl}).
$
The first one, 
$-98\, (1) \!\times\! 10^{-11}$\cite{HMNT06},
is the $O(\alpha^3)$ contribution of diagrams containing hadronic
vacuum polarization insertions~\cite{Kr96}. The second term, also of
$O(\alpha^3)$, is the hadronic light-by-light contribution; as it
cannot be determined from data, its evaluation relies on specific
models. Recent determinations of this term vary between
$80(40) \!\times\! 10^{-11}$\cite{Andreas}
and
$136(25) \!\times\! 10^{-11}$\cite{Arkady}.
The most recent one,
$110(40) \!\times\! 10^{-11}$\cite{BP07},
lies between them. If we add this result to 
$ a_{\mu}^{\mbox{$\scriptscriptstyle{\rm HLO}$}}$, for example the value of Ref.~\cite{HMNT06} (which also provides the hadronic contribution to the effective fine-structure constant, later required for our discussion), and the rest of the {\small SM} contributions, we obtain
$     \amu^{\mbox{$\scriptscriptstyle{\rm SM}$}}= 116591778(61)  
\!\times\! 10^{-11}$.
The difference with the experimental value
$
    a_{\mu}^{\mbox{$\scriptscriptstyle{\rm EXP}$}}  =
               116592080(63) \!\times\! 10^{-11}
$~\cite{bnl}
is 
$\Delta a_{\mu} = a_{\mu}^{\mbox{$\scriptscriptstyle{\rm EXP}$}}-
\amu^{\mbox{$\scriptscriptstyle{\rm SM}$}} = +302(88) \!\times\! 10^{-11}$,
i.e., 3.4$\sigma$ (all errors were added in quadrature).  Similar discrepancies are found employing the 
$a_{\mu}^{\mbox{$\scriptscriptstyle{\rm HLO}$}}$ values reported in Refs.~\cite{DE07,Jeger06,TY05}. For recent reviews of $\amu$ see~\cite{TeubnerSUSY08,Reviews,DM04}.

The term $a_{\mu}^{\mbox{$\scriptscriptstyle{\rm HLO}$}}$ can alternatively be computed incorporating hadronic $\tau$-decay data, related to those of hadroproduction in $e^+e^-$ collisions via isospin symmetry~\cite{ADH98,DEHZ}. Unfortunately there is a large difference between the $e^+e^-$- and $\tau$-based determinations of $a_{\mu}^{\mbox{$\scriptscriptstyle{\rm HLO}$}}$, even if isospin violation corrections are taken into account~\cite{IVC1}. The $\tau$-based value is significantly higher, leading to a small ($\sim 1 \sigma$) $\Delta a_{\mu}$ difference. As the $e^+e^-$ data are more directly related to the $\amu^{\mbox{$\scriptscriptstyle{\rm HLO}$}}$ calculation than the $\tau$ ones, all recent analyses do not include the latter. Also, we note that recently studied additional isospin-breaking corrections somewhat reduce the difference between these two sets of data (lowering the $\tau$-based determination)~\cite{IVC2,IVC3}, and a new analysis of the pion form factor claims that the $\tau$ and $e^+e^-$ data are consistent after isospin violation effects and vector meson mixings are considered~\cite{IVC4}.

The 3.4$\sigma$ discrepancy between the theoretical prediction and the experimental value of the muon \gmt can be explained in several ways. It could be due, at least in part, to an error in the determination of the hadronic light-by-light contribution. However, if this were the only cause of the discrepancy, $a_{\mu}^{\mysmall \rm HHO}(\mbox{lbl})$ would have to move up by many standard deviations to fix it --~roughly eight, if we use the $a_{\mu}^{\mysmall \rm HHO}(\mbox{lbl})$ result of Ref.~\cite{BP07} (which includes all known uncertainties), and more than ten if the estimate of Ref.~\cite{Arkady} is employed instead.  Although the errors assigned to $a_{\mu}^{\mysmall \rm HHO}(\mbox{lbl})$ are only educated guesses, this solution seems unlikely, at least as the dominant one.

Another possibility is to explain the discrepancy $\Delta a_{\mu}$ via the {\small   QED}, {\small EW} and hadronic higher-order vacuum polarization contributions; this looks very improbable, as one can immediately conclude inspecting their values and uncertainties reported above. If we assume that the \gmt experiment {\small   E821} is correct, we are left with two options: possible contributions of physics beyond the {\small SM}, or an erroneous determination of the leading-order hadronic contribution $a_{\mu}^{\mbox{$\scriptscriptstyle{\rm HLO}$}}$ (or both). The first of these two explanations has been extensively discussed in the literature; following Ref.~\cite{Passera:2008jk} we will study whether the second one is realistic or not, and analyze its implications for the {\small EW} bounds on the mass of the Higgs boson.

\section{Errors in the hadronic cross section?}

The hadronic leading-order contribution 
$\amu^{\mbox{$\scriptscriptstyle{\rm HLO}$}}$ can be computed via the dispersion integral~\cite{DISPamu}
\be
      a_{\mu}^{\mbox{$\scriptscriptstyle{\rm HLO}$}}= 
      \frac{1}{4\pi^3} \!
      \int^{\infty}_{4m_{\pi}^2} ds \, K(s) \, \sigma (s),
\label{eq:amudispint}
\ee
where $\sigma (s)$ is the total cross section for $e^+ e^-$ annihilation into any hadronic state, with extraneous {\small QED} corrections subtracted off, and $s$ is the squared momentum transfer. The well-known kernel function $K(s)$ (see~\cite{EJ95}) is positive definite, decreases monotonically for increasing $s$ and, for large $s$, behaves as $m_\mu^2/(3s)$ to a good approximation. About 90\% of the total contribution to $\amu^{\mbox{$\scriptscriptstyle{\rm HLO}$}}$ is accumulated at center-of-mass energies $\sqrt{s}$ below 1.8~GeV and roughly three-fourths of $\amu^{\mbox{$\scriptscriptstyle{\rm HLO}$}}$ is covered by the two-pion final state which is dominated by the $\rho(770)$ resonance~\cite{DEHZ}. Exclusive low-energy $e^+e^-$ cross sections were measured at colliders in Frascati, Novosibirsk, Orsay, and Stanford, while at higher energies the total cross section was determined inclusively.

Let's now assume that the discrepancy
$\Delta a_{\mu} = a_{\mu}^{\mbox{$\scriptscriptstyle{\rm EXP}$}}-
\amu^{\mbox{$\scriptscriptstyle{\rm SM}$}} = +302(88) \!\times\! 10^{-11}$,
is due to --~and only to~-- hypothetical errors in $\sigma (s)$, and let us increase this cross section in order to raise 
$\amu^{\mbox{$\scriptscriptstyle{\rm  HLO}$}}$, thus reducing $\Delta a_{\mu}$. This simple assumption leads to interesting consequences. An upward shift of the hadronic cross section also induces an increase of the value of the hadronic contribution to the effective fine-structure constant at $M_Z$~\cite{DISPDalpha},
\be
 \dafive = \frac{\mz^2}{4 \alpha \pi^2}
  \,\, P \! \int_{4m_\pi^2}^{\infty} ds \, \frac{\sigma(s)}{\mz^2 -s}
\label{eq:Dpi5dispint}
\ee       
($P$ stands for Cauchy's principal value).  This integral is similar to the one we encountered in \eq{amudispint} for $a_{\mu}^{\mbox{$\scriptscriptstyle{\rm HLO}$}}$. There, however, the weight function in the integrand gives a stronger weight to low-energy data. 
Let us define
$
     a_i = \int_{4m_{\pi}^2}^{s_u}ds \, f_i(s) \, \sigma (s)
$
$(i=1,2)$, where the upper limit of integration is $s_u < \mz^2$, and the kernels are $f_1(s) = K(s)/(4 \pi^3)$ and $f_2(s) = [\mz^2/(\mz^2-s)]/(4 \alpha \pi^2)$. The integrals $a_i$ with $i=1,2$ provide the contributions to $a_{\mu}^{\mbox{$\scriptscriptstyle{\rm HLO}$}}$ and $\dafive$, respectively,
from $4m_\pi^2$ up to $s_u$ (see Eqs.~(\ref{eq:amudispint},\ref{eq:Dpi5dispint})).
An increase of the cross section $\sigma(s)$ of the form
$
     \Delta \sigma(s) = \epsilon \sigma(s)
$
in the energy range $\sqrt s \in [\sqrt s_0 - \delta/2, \sqrt s_0 + \delta/2]$, where $\epsilon$ is a positive constant and $2m_{\pi}+\delta/2<\sqrt s_0<\sqrt s_u -\delta/2$, increases $a_1$ by $\Delta a_1 (\sqrt s_0,\delta,\epsilon) = \epsilon \int_{\sqrt   s_0-\delta/2}^{\sqrt s_0+\delta/2} 2t \, \sigma(t^2) \, f(t^2) \, dt$. If we assume that the muon \gmt discrepancy is entirely due to this increase in $\sigma(s)$, so that $\Delta a_1 (\sqrt s_0,\delta,\epsilon) = \Delta a_{\mu}$, the corresponding increase in $\dafive$ is
\be 
\Delta a_2(\sqrt s_0,\delta) = \Delta a_{\mu} 
\frac{\int_{\sqrt s_0-\delta/2}^{\sqrt s_0+\delta/2} g(t^2) \, 
  \sigma(t^2) \, t \, dt} 
     {\int_{\sqrt s_0-\delta/2}^{\sqrt s_0+\delta/2} f(t^2) \, 
  \sigma(t^2) \, t \, dt}.
\label{eq:shiftb} 
\ee
The shifts $\Delta a_2(\sqrt s_0,\delta)$ were studied in
Ref.~\cite{Passera:2008jk} for several bin widths $\delta$ and central
values $\sqrt s_0$.


The present global fit of the {\small LEP} Electroweak Working Group ({\small   EWWG}) leads to the Higgs boson mass
$\mh \!=\! 84^{+34}_{-26}$~GeV 
and the 95\% confidence level ({\small CL}) upper bound $\mhUB \!\simeq\! 154$~GeV~\cite{newLEPEWWG}. This result is based on the recent preliminary top quark mass $\mt\!=\!172.4(1.2)$~GeV~\cite{Group:2008vn} and the value
$\dafive \!=\! 0.02758(35)$~\cite{BP05}.
The {\small LEP} direct-search 95\%{\small CL} lower bound is $\mhLB=114.4$~GeV~\cite{MHLB03}.
Although the global {\small EW} fit employs a large set of observables, $\mhUB$ is strongly driven by the comparison of the theoretical predictions of the W boson mass and the effective {\small EW} mixing angle $\seff$ with their precisely measured values. Convenient formulae providing the $\mw$ and $\seff$ {\small SM} predictions in terms of $\mh$, $\mt$, $\dafive$, and $\alpha_s(\mz)$, the strong coupling constant at the scale $\mz$, are given in~\cite{formulette}.  Combining these two predictions via a numerical $\chi^2$-analysis and using the present world-average values
$\mw \!=\! 80.399(25)$~GeV~\cite{Wmass}, 
$\seff \!=\! 0.23153(16)$~\cite{LEPEWWG05},
$\mt \!=\! 172.4(1.2)$~GeV~\cite{Group:2008vn}, 
$\alpha_s(\mz) \!=\! 0.118(2)$~\cite{PDG08},
and the determination
$\dafive = 0.02758(35)$~\cite{BP05},
we get
$\mh = 89^{+37}_{-27}$~GeV
and $\mhUB=156$~GeV. We see that indeed the $\mh$ values obtained from
the $\mw$ and $\seff$ predictions are quite close to the results of
the global analysis.

The $\mh$ dependence of $\amu^{\mbox{$\scriptscriptstyle{\rm SM}$}}$ is too weak to provide $\mh$ bounds from the comparison with the measured value.
On the other hand, $\dafive$ is one of the key inputs of the {\small EW} fits. For example, employing the more recent (and slightly higher) value
$\dafive = 0.02768(22)$~\cite{HMNT06}
instead of 0.02758(35)~\cite{BP05}, the $\mh$ prediction shifts down
to
$\mh = 88^{+32}_{-24}$~GeV 
and $\mhUB=145$~GeV. In~\cite{Passera:2008jk} we considered the new values of $\dafive$ obtained shifting 0.02768(22)~\cite{HMNT06} by $\Delta a_2(\sqrt s_0, \delta)$ (including their uncertainties), and computed the corresponding new values of $\mhUB$ via the combined $\chi^2$-analysis based on the $\mw$ and $\seff$ inputs (for both $\dafive$ and $\amu^{\mbox{$\scriptscriptstyle{\rm HLO}$}}$ we used the values reported in~\cite{HMNT06}). Our results show that an increase
$\epsilon \sigma (s)$
of the hadronic cross section (in $\sqrt s \in [\sqrt s_0 - \delta/2, \sqrt s_0 + \delta/2]$), adjusted to bridge the muon \gmt discrepancy $\Delta a_{\mu}$, decreases $\mhUB$, further restricting the already narrow allowed region for $\mh$. We concluded that these hypothetical shifts conflict with the lower limit $\mhLB$ when
$\sqrt s_0 \gtrsim 1.2$~GeV, 
for values of $\delta$ up to several hundreds of MeV. In~\cite{Passera:2008jk} we pointed out that there are more complex scenarios where it is possible to bridge the $\Delta a_{\mu}$ discrepancy without significantly affecting $\mhUB$, but they are considerably more unlikely than those discussed above.

If $\tau$ data are incorporated in the calculation of the dispersive integrals in Eqs.~(\ref{eq:amudispint},\ref{eq:Dpi5dispint}), $a_{\mu}^{\mbox{$\scriptscriptstyle{\rm HLO}$}}$ significantly increases to
$7110(58) \!\times\! 10^{-11}$\cite{DEHZ}, 
$\amu^{\mbox{$\scriptscriptstyle{\rm HHO}$}}(\mbox{vp})$ 
slightly decreases to      
$-101(1)\!\times\! 10^{-11}$\cite{HMNT06,DM04}, 
and the discrepancy drops to
$\Delta a_{\mu} = +89(95) \!\times\! 10^{-11}$,
i.e.\ $\sim 1\sigma$. While using $\tau$ data almost solves the $\Delta a_{\mu}$ discrepancy, it increases $\dafive$ to 0.02782(16)~\cite{Marciano04,DEHZ}. In~\cite{Marciano04} it was shown that this increase leads to a low $\mh$ prediction which is suggestive of a near conflict with $\mhLB$, leaving a narrow window for $\mh$.  Indeed, with this value of $\dafive$ and the same above-discussed other inputs of the $\chi^2$-analysis, we find an $\mhUB$ value of only 133~GeV.

Recently computed isospin-breaking violations, improvements of the long-distance radiative corrections to the decay $\tau^- \to \pi^- \pi^0 \nu_{\tau}$~\cite{IVC2}, and differentiation of the neutral and charged $\rho$ properties~\cite{IVC3} reduce to some extent the difference between $\tau$ and $e^+e^-$ data, lowering the $\tau$-based determination of $ a_{\mu}^{\mbox{$\scriptscriptstyle{\rm HLO}$}}$. Moreover, a recent analysis of the pion form factor below 1~GeV claims that $\tau$ data are consistent with the $e^+e^-$ ones after isospin violation effects and vector meson mixings are considered~\cite{IVC4}. In this case one could use the $e^+e^-$ data below $\sim \!1$~GeV, confirmed by the $\tau$ ones, and assume that $\Delta a_{\mu}$ is accommodated by hypothetical errors occurring above $\sim \!1$~GeV, where disagreement persists between these two data sets. Reference~\cite{Passera:2008jk} shows that this assumption would lead to $\mhUB$ values inconsistent with $\mhLB$.


In the above analysis, the hadronic cross section $\sigma(s)$ was shifted up by amounts $\Delta \sigma(s) = \epsilon \sigma(s)$ adjusted to bridge $\Delta a_{\mu}$. Apart from the implications for $\mh$, these shifts may actually be inadmissibly large when compared with the quoted experimental uncertainties. Consider the parameter $\epsilon=\Delta \sigma(s)/\sigma(s)$. Clearly, its value depends on the choice of the energy range $[\sqrt s_0 - \delta/2, \sqrt s_0 + \delta/2]$ where $\sigma(s)$ is increased and, for fixed $\sqrt s_0$, it decreases when $\delta$ increases. Its minimum value, $\sim 4\%$, occurs if $\sigma(s)$ is multiplied by $(1+\epsilon)$ in the whole integration region, from $2\mpi$ to infinity. Such a shift would lead to $\mhUB \sim 70$~GeV, well below $\mhLB$.  Higher values of $\epsilon$ are obtained for narrower energy bins, particularly if they do not include the $\rho$-$\omega$ resonance region. For example, a huge $\epsilon \sim 52\%$ increase is needed to accommodate $\Delta a_{\mu}$ with a shift of $\sigma(s)$ in the region from $2\mpi$ up to 500~MeV (reducing $\mhUB$ to 139~GeV), while an increase in a bin of the same size but centered at the $\rho$ peak requires $\epsilon \sim 8\%$ (lowering $\mhUB$ to 127~GeV). As the quoted experimental uncertainty of $\sigma(s)$ below 1~GeV is of the order of a few per cent (or less, in some specific energy regions), the possibility to explain $\Delta a_{\mu}$ with these shifts $\Delta \sigma(s)$ appears to be unlikely. Lower values of $\epsilon$ are obtained if the shifts occur in energy ranges centered around the $\rho$-$\omega$ resonances, but also this possibility looks unlikely, since it requires variations of $\sigma(s)$ of at least $\sim 6$\%. If, however, such shifts $\Delta \sigma(s)$ indeed turn out to be the solution of the $\Delta a_{\mu}$ discrepancy, then $\mhUB$ is reduced to about 130~GeV~\cite{Passera:2008jk}.

It is interesting to note that in the scenario where $\Delta a_{\mu}$ is due to hypothetical errors in $\sigma(s)$, rather than ``new physics'', the reduced $\mhUB \lesssim 130$~GeV induces some tension with the approximate 95\% {\small CL} lower bound $\mh \gtrsim 120$~GeV required to ensure vacuum stability under the assumption that the {\small SM} is valid up to the Planck scale~\cite{VacuumStability} (note, however, that this lower bound somewhat decreases when the vacuum is allowed to be metastable, provided its lifetime is longer than the age of the universe~\cite{VacuumMetaStability}). Thus, one could argue that this tension is, on its own, suggestive of physics beyond the {\small SM}.

We remind the reader that the present values of $\seff$ derived from the leptonic and hadronic observables are respectively $(\seff)_{l} \! = \! 0.23113(21)$ and $(\seff)_{h} \! = \! 0.23222(27)$~\cite{LEPEWWG05}. In Ref.~\cite{Passera:2008jk} we pointed out that the use of either of these values as an input parameter leads to inconsistencies in the {\small SM} framework that already require the presence of ``new physics''. For this reason, we followed the standard practice of employing as input the world-average value for $\seff$ determined in the {\small SM} global analysis. Since $\mhUB$ also depends sensitively on $\mt$, in~\cite{Passera:2008jk} we provide simple formulae to obtain the new values derived from different $\mt$ inputs.

\section{Conclusions}

We examined a number of hypothetical errors in the {\small SM} prediction of the muon \gmt that could be responsible for the present discrepancy $\Delta a_{\mu}$ with the experimental value. None of them looks likely.  In particular, following Ref.~\cite{Passera:2008jk} we showed how an increase $\Delta \sigma(s)\!=\!\epsilon \sigma(s)$ of the hadroproduction cross section in low-energy $e^+e^-$ collisions could bridge $\Delta a_{\mu}$. However, such increases lead to reduced $\mh$ upper bounds (lower than 114.4~GeV -- the {\small LEP} lower bound -- if they occur in energy regions centered above $\sim 1.2$~GeV). Moreover, their amounts are generally very large when compared with the quoted experimental uncertainties, even if the latter were significantly underestimated. The possibility to bridge the muon \gmt discrepancy with shifts of the hadronic cross section therefore appears to be unlikely. If, nonetheless, this turns out to be the solution, then the 95\% {\small CL} upper bound $\mhUB$ drops to about 130~GeV.

If $\tau$-decay data are incorporated in the calculation of $a_{\mu}^{\mysmall \rm SM}$, the muon \gmt discrepancy decreases to $\sim \!1 \sigma$. While this almost solves $\Delta a_{\mu}$, it raises the value of $\dafive$ leading to $\mhUB= 133$~GeV, increasing the tension with the {\small LEP} lower bound. One could also consider a scenario, suggested by recent studies, where the $\tau$ data confirm the $e^+e^-$ ones below $\sim \! 1$~GeV, while a discrepancy between them persists at higher energies. If, in this case, $\Delta a_{\mu}$ is fixed by hypothetical errors above $\sim \! 1$~GeV, where the data sets disagree, one also finds values of $\mhUB$ inconsistent with the {\small LEP} lower bound.

It has been suggested~\cite{Schiel:2007aw} that a $P$-wave electromagnetic bound state of $\pi^+ \pi^-$ (pionium) could enter the dispersion relations through 1\% mixing with the $\rho$ in a way that significantly increases the hadronic contribution to $\amu$. If so, such a state would give little change to the Higgs mass determination and would seem to refute our claims.  However, Ref.~\cite{Schiel:2007aw} is in error. The required mixing is actually 0.1, not the erroneous 0.01 claim in \cite{Schiel:2007aw}, and such large mixing is not possible.  The actual effect of pionium on $\amu$ is negligible~\cite{Vainshtein}.

If the $\Delta a_{\mu}$ discrepancy is real, it points to ``new physics'', like low-energy supersymmetry where $\Delta a_{\mu}$ is reconciled by the additional contributions of supersymmetric partners and one expects $\mh \! \lesssim \! 135$~GeV for the mass of the lightest scalar~\cite{DHHSW}.  If, instead, the deviation is caused by an incorrect leading-order hadronic contribution, it leads to reduced $\mhUB$ values. This reduction, together with the {\small LEP} lower bound, leaves a narrow window for the mass of this fundamental particle. Interestingly, it also raises the tension with the $\mh$ lower bound derived in the {\small SM} from the vacuum stability requirement.

\vspace{2mm}
\noindent {\bf Acknowledgments.} We thank A.\ Vainshtein for discussions and correspondence. The work of {\small W.J.M.}, {\small M.P.}, and {\small A.S.} was supported by {\small U.S.\ DOE} grant {\small DE-AC02-76CH00016}, {\small   E.C.} contracts {\small MRTN-CT 2004-503369 \& 2006-035505}, and {\small   U.S.\ NSF} grant {\small PHY-0758032}, respectively.

\vspace{-2mm}


\begin{thebibliography}{99}

\bibitem{bnl} G.W.~Bennett {\it et al.}, 
  Phys.\ Rev.\ D {\bf 73} (2006) 072003; 
  Phys.\ Rev.\ Lett.\ {\bf 92} (2004) 161802; 
                      {\bf 89} (2002) 101804; 
                      {\bf 89} (2002) 129903(E).

\bibitem{NP}
  A.~Czarnecki and W.J.~Marciano,
  Phys.\ Rev.\ D {\bf 64} (2001) 013014;
  D.~St\"{o}ckinger, J.\ Phys.\ G {\bf 34} (2007) R45; 
  D.~Nomura, these proceedings.

\bibitem{QED}
  T.~Kinoshita, M.~Nio,
  Phys.\ Rev.\ D {\bf 73} (2006) 013003;
                 {\bf 70} (2004) 113001;
                 {\bf 73} (2006) 053007;
  T.~Aoyama {\it et al.},
  Phys.\ Rev.\ Lett.\ {\bf 99} (2007) 110406;
  Phys.\ Rev.\  D{\bf77} (2008) 053012;
  S.~Laporta, E.~Remiddi,
  Phys.\ Lett.\  B {\bf 301} (1993) 440;
                   {\bf 379} (1996) 283;
  M.~Passera,  Phys.\ Rev.\  D {\bf 75} (2007) 013002;
  A.L.~Kataev, Phys.\ Rev.\  D {\bf 74} (2006) 073011.

\bibitem{EW}
  A.~Czarnecki, W.J.~Marciano and A.~Vainshtein,
  Phys.\ Rev.\  D {\bf 67} (2003) 073006;
                D {\bf 73} (2006) 119901(E);
  A.~Czarnecki, B.~Krause and W.J.~Marciano,
  Phys.\ Rev.\  D {\bf 52} (1995) 2619;
  Phys.\ Rev.\ Lett.\  {\bf 76} (1996) 3267.

\bibitem{DE07}
  M.~Davier,
  Nucl.\ Phys.\ Proc.\ Suppl.\ {\bf 169} (2007) 288;
  S.~Eidelman,
  Acta Phys.\ Polon.\ B {\bf 38} (2007) 3015.

\bibitem{HMNT06}
  K.~Hagiwara {\it et al.}, 
  Phys.\ Lett.\  B {\bf 649} (2007) 173.

\bibitem{TeubnerSUSY08}
T.~Teubner, these proceedings.

\bibitem{Jeger06}
  F.~Jegerlehner,
  Nucl.\ Phys.\ Proc.\ Suppl.\  {\bf 162} (2006) 22.

\bibitem{TY05}
  J.F.de Troconiz,F.J.Yndurain,~Phys.Rev.D{\bf71}(2005)73008.

\bibitem{Kr96}
  B.~Krause, Phys.\ Lett.\  B {\bf 390} (1997) 392.

\bibitem{Andreas}
  M.\ Knecht, A.\ Nyffeler, Phys.\ Rev.\ D {\bf 65} (2002) 73034;
  M.~Knecht {\it et al.},
  Phys.\ Rev.\ Lett.\ {\bf 88} (2002) 71802.

 \bibitem{Arkady}
  K.Melnikov, A.Vainshtein,
  Phys.Rev.D{\bf70} (2004) 113006.

\bibitem{BP07}
  J.~Bijnens, J.~Prades,
  Mod.\ Phys.\ Lett.\  A {\bf 22} (2007) 767.

\bibitem{Reviews}
  F.~Jegerlehner,
  Acta Phys.\ Polon.\ B {\bf 38} (2007) 3021;
  {\it The anomalous magnetic moment of the muon}, Springer 2007;
  J.P.~Miller, E.~de Rafael and B.L.~Roberts,
  Rept.\ Prog.\ Phys.\  {\bf 70} (2007) 795;
  M.~Passera,
  Nucl.\ Phys.\ Proc.\ Suppl.\  {\bf 169} (2007) 213;
                                {\bf 162} (2006) 242;
                                {\bf 155} (2006) 365;
   J.\ Phys.\ G {\bf 31} (2005) R75;
  K.~Melnikov and A.~Vainshtein, 
  {\it Theory of the muon anomalous magnetic moment}, Springer 2006.

\bibitem{DM04}
  M.Davier,W.J.Marciano, Ann.Rev.Nucl.Sci.{\bf54}(2004)115.

\bibitem{ADH98} R.Alemany,M.Davier,A.H\"ocker,Eur.Phys.J.C{\bf2}(1998)123.

\bibitem{DEHZ}
  M.Davier {\it et al.},
  Eur.Phys.J.C{\bf27}(2003)497;C{\bf31}(2003)503.

\bibitem{IVC1}
  W.J.~Marciano, A.~Sirlin,
  Phys.\ Rev.\ Lett.\ {\bf 61} (1988) 1815;
  A.\ Sirlin, Nucl.\ Phys.\  B{\bf196} (1982) 83;
  V.~Cirigliano {\it et al.},
  Phys.\ Lett.\ B {\bf 513} (2001) 361;
  {\small JHEP} {\bf 0208} (2002) 002.

\bibitem{IVC2} 
  F.~Flores-Baez {\it et al.}, Phys.\ Rev.\ D {\bf 74} (2006) 071301.

\bibitem{IVC3} 
  F.V.~Flores-Baez {\it et al.}, Phys.\ Rev.\ D {\bf 76} (2007) 096010.

\bibitem{IVC4}
  M.~Benayoun {\it et al.},
  Eur.\ Phys.\ J.\  C {\bf 55} (2008) 199.

\bibitem{Passera:2008jk}
  M.~Passera, W.J.~Marciano and A.~Sirlin,
  Phys.\ Rev.\  D {\bf 78} (2008) 013009.

\bibitem{DISPamu}
  C.~Bouchiat, L.\ Michel, J.Phys.Radium 22 (1961) 121;
  M.~Gourdin, E.~de Rafael, Nucl.\ Phys.\ B {\bf 10} (1969) 667.

\bibitem{EJ95}
  S.~Eidelman and F.~Jegerlehner,
  Z.\ Phys.\  C {\bf 67} (1995) 585.

\bibitem{DISPDalpha}
  N.~Cabibbo and R.~Gatto, Phys.\ Rev.\ {\bf 124} (1961) 1577.

\bibitem{newLEPEWWG} LEP EW Working Group, 
http://lepewwg.web.cern.ch.

\bibitem{Group:2008vn} 
  Tevatron EW Working Group, 
  arXiv:0808.1089 [hep-ex].

\bibitem{BP05}
  H.~Burkhardt, B.~Pietrzyk,
  Phys.\ Rev.\ D{\bf72} (2005) 057501.

\bibitem{MHLB03}
  R.~Barate {\it et al.},
  Phys.\ Lett.\  B {\bf 565} (2003) 61.

\bibitem{formulette}
  G.Degrassi {\it et al.}, 
  Phys.\ Lett.\ B{\bf418}(1998)209;
  G.Degrassi, P.Gambino,
  Nucl.\ Phys.\ B{\bf567}(2000)3;
  A.~Ferroglia {\it et al.}, 
  Phys.Rev.D{\bf65}(2002)113002;
  M.Awramik {\it et al.},Phys.Rev. D{\bf69} (2004) 053006;
  Phys.\ Rev.\ Lett.\  {\bf 93} (2004) 201805;
  M.Awramik, M.Czakon, A.Freitas,
  JHEP{\bf0611}(2006)048.

\bibitem{Wmass}
  J.~Alcaraz {\it et al.}, arXiv:hep-ex/0612034;
  Tevatron EW Working Group, 
  arXiv:0808.0147 [hep-ex].

\bibitem{LEPEWWG05} 
  M.~Gr\"{u}newald  {\it et al.}, Phys.\ Rept.\ {\bf 427} (2006) 257.

\bibitem{PDG08}
  C.~Amsler {\it et al.}  [PDG],
  Phys.\ Lett.\  B {\bf 667} (2008) 1.

\bibitem{Marciano04}
  W.J.~Marciano, arXiv:hep-ph/0411179.

\bibitem{VacuumStability}
  G.Altarelli, G.Isidori,
  Phys.\ Lett.\  B{\bf337} (1994) 141;
  J.A.\ Casas,  {\it et al.},
   Phys.Lett.B{\bf342}(1995)171;
             B{\bf382}(1996)374.

\bibitem{VacuumMetaStability}
  G.~Isidori {\it et al.}, Nucl.\ Phys.\ B {\bf 609} (2001) 387.

\bibitem{Schiel:2007aw}
  R.W.~Schiel and J.P.~Ralston, Phys.\ Lett.\  B {\bf 657} (2007) 43.

\bibitem{Vainshtein}  A.~Vainshtein, private communication. Prof.\ Vainshtein reaches the same conclusions as us. 

\bibitem{DHHSW}
  G.~Degrassi {\it et al.},
  Eur.\ Phys.\ J.\  C {\bf 28} (2003) 133.



\end{thebibliography}
\end{document}